\documentclass[9pt]{article}

\usepackage{graphicx}
\usepackage{float}
\usepackage{url}
\usepackage{authblk}

\title{\textbf{DSiD: a Delphes Detector for ILC Physics Studies} \\  \vspace{0.2in} \large  \textit{Talk presented at the International Workshop on Future Linear Colliders (LCWS15), Whistler, Canada, 2-6 November 2015}}

\author{C.T. Potter}
\affil{University of Oregon}
\date{\today}

\begin{document}

\maketitle

\abstract{We describe DSiD, a fast simulation Delphes detector for the International Linear Collider (ILC) based on the full simulation performance of the SiD detector. SiD is one of two detectors described in the ILC Technical Design Report (TDR). The tracking efficiency, tracking momentum resolution, electromagnetic and hadronic calorimeter energy resolution, particle identification and flavor tagging efficiencies are taken from the Detailed Baseline Design (DBD) study as described in ILC TDR Volume 4: Detectors. In a cross-check study with $4 \times 10^6$ $e^+ e^- \rightarrow b \bar{b}$ events generated at $\sqrt{s}=500$~GeV and simulated by Delphes with the DSiD detector card, these performance characteristics are measured and found to be commensurate with the DBD results. For a new physics use case example, we describe a study of Next-to-Minimal Supersymmetric $e^+ e^- \rightarrow 2 \chi_3 \rightarrow 2\chi_1 2 h_1$ with $h_1 \rightarrow 2a_1$ or $h_1 \rightarrow b\bar{b}$ at the $\sqrt{s}=500$~GeV ILC. The card is available on HepForge.}

\section{Introduction}

With the discovery of the Standard Model Higgs boson at the Large Hadron Collider (LHC) \cite{Aad:2012tfa,Chatrchyan:2012ufa}, the scientific case for the International Linear Collider (ILC) is now well established \cite{Baer:2013cma,Dawson:2013bba,Asner:2013psa}. But the LHC has only just begun to explore the energy reach of the ILC, where the shortcomings of the SM tell us there must be new physics. 

The machinery for full detector simulation of ILC events is well established \cite{Graf:2006ei,1742-6596-513-2-022010}, and some fast simulation machinery is in place \cite{Berggren:2012ar}. If, as expected, the LHC experiments announce discovery of new physics in Run 2, the need for rapid ILC phenomenology studies to evaluate the capability of the ILC for measuring the new physics will be evident. In this brief report we describe a Delphes \cite{deFavereau:2013fsa,Selvaggi:2014mya,Mertens:2015kba,Cacciari:2011ma} fast detector simulation card modeled on the full simulation SiD detector performance as described in the ILC Technical Design Report (TDR) Volume 4: Detectors \cite{Behnke:2013lya}.

SiD is designed to provide excellent momentum and energy resolution over the broad range of particle energies expected at the ILC. It features a 5T solenoidal magnetic field, a vertex detector instrumented with silicon pixels for vertex reconstruction and a main tracker instrumented with silicon strips for measuring charged particle momentum. The electromagnetic calorimeter uses silicon  strips in the active layers and tungsten in the passive layers for measuring electromagnetic energy deposits. The hadronic calorimeter employs glass resistive plate chambers in the active layers and steel in the passive layers for measuring hadronic energy deposits. The muon system is instrumented with scintillators in the iron flux return. Full details of the SiD design can be found in \cite{Behnke:2013lya}. 

Delphes, a multipurpose fast detector simulator which can read StdHep \cite{stdhep}, LHEF \cite{Alwall:2006yp} and HepMC \cite{Dobbs:2001ck} event formats, has been used extensively for LHC $pp$ phenomenology studies. But nothing in its architecture prevents it from use in an ILC $e^+ e^-$ environment.  The current Delphes distribution now includes a detector card for the ILC Large Detector (ILD), the other ILC detector presented in the ILC TDR. We describe here the DSiD card available on HepForge \cite{Buckley:2006nm} at \texttt{dsid.hepforge.org}.

\begin{table}[p]
\begin{center}
\begin{tabular}{|c|c|} \hline
Parameter & Value \\ \hline
\multicolumn{2}{|c|}{ParticlePropagator} \\ \hline
 Radius  & 2.493m \\
 HalfLength  & 3.018m \\
 Bz & 5.0 T \\ \hline
\multicolumn{2}{|c|}{TrackingEfficiency} \\ \hline
ChargedHadronTrackingEfficiency &  see DBD Figure 3.5\\
ElectronTrackingEfficiency &  see DBD Figure 3.5\\
MuonTrackingEfficiency &  see DBD Figure 3.5\\ \hline
\multicolumn{2}{|c|}{MomentumSmearing} \\ \hline
ChargedHadronMomentumSmearing &  see DBD Figure 3.9\\ 
MuonMomentumSmearing &   see DBD Figure 3.9\\ 
ElectronEnergySmearing &  see DBD Figure 3.9 \\ \hline
\multicolumn{2}{|c|}{ECal,HCal} \\ \hline
ECal ResolutionFormula  & $\sigma_{E}/E = 0.010 \oplus 0.170/\sqrt{E}$\\
HCal ResolutionFormula & $\sigma_{E}/E = 0.094 \oplus 0.559/\sqrt{E}$\\ \hline
\multicolumn{2}{|c|}{Photon,Electron,Muon Efficiency} \\ \hline
PhotonEfficiency &  see DBD Figure 10.6 \\ 
ElectronEfficiency &   see DBD Figure 10.7\\
MuonEfficiency &   see DBD Figure 10.8 \\ \hline
\multicolumn{2}{|c|}{FastJetFinder} \\ \hline
JetAlgorithm & 6 [anti$k_t$]\\
ParameterR & 1.0\\
InputArray & EFlowMerger/eflow\\ \hline
\multicolumn{2}{|c|}{BTagging} \\ \hline
EfficiencyFormula {0} & 0.007  \\
EfficiencyFormula {4} &  0.03\\
EfficiencyFormula {5} &  0.7 \\ \hline
\multicolumn{2}{|c|}{TauTagging} \\ \hline
EfficiencyFormula {0} & 0.001 \\
EfficiencyFormula {15} &  0.4\\ \hline
\end{tabular}
\caption{Delphes modules and their parameter values in DSiD. The Figures refer to the DBD figures used to specify the DSiD performance, which are reproduced in the Figures of this report for comparison to the performance measured in this study.} 
\label{tab:card}
\end{center}
\end{table}

\section{Performance Specification}

Delphes simulates detector response using efficiencies and resolutions parameterized by a particle's transverse momentum and pseudorapidity or, equivalently, polar angle with respect to the beam axis. The modules which simulate the tracking response to electrons, muons and charged hadrons, in order of execution, are below.

\begin{itemize}

\item ParticlePropagator propagates all stable particles through the specified magnetic field

\item TrackingEfficiency either kills the particle or not based on the parameterized efficiencies

\item MomentumSmearing smears the particle momentum according to the parameterized resolution

\item TrackMerger merges electron, muon and charged hadron tracks

\end{itemize}

For the Delphes calorimetery, either a single generic calorimeter can be specified for both electromagnetic and hadronic calorimetry or, as for DSiD, two separate calorimeters can be specified. For the latter case, the modules, in order of execution, are below.

\begin{itemize}

\item TrackMerger supplies track information to the calorimetry

\item ECal separates eflowTracks from eflowPhotons and applies energy smearing

\item HCal separates eflowTracks from eflowNeutralHadrons and applies energy smearing

\item EFlowMerger merges eflowTracks with eflowNeutralHadrons and eflowPhotons

\end{itemize}

\noindent The EFlowMerger is used in case the jetfinding uses energy flow objects rather than calorimeter towers, as is the case for SiD and ILD \cite{Thomson200925}. Energy (or particle) flow algorithms associate tracks to calorimeter deposits to capitalize on superior track momentum resolution, thus giving the best possible jet energy resolution. Six possible jetfinding algorithms can be specified in the FastJetFinder module; for DSiD it is the anti$k_t$ algorithm.

Particle identification efficiencies are specified in the ElectronEfficiency, PhotonEfficiency, MuonEfficiency, BTagging and TauTagging modules. Relative or absolute isolation can be specified in the ElectronIsolation, PhotonIsolation and MuonIsolation modules. 

Part II of \cite{Behnke:2013lya}, the SiD Detailed Baseline Design (DBD), contains enough information to completely specify a Delphes card. See Table \ref{tab:card} for the parameters used for the DSiD card. The tracking efficiencies for muons, electrons and charged pions are taken from Figure 3.5 of the DBD, the momentum smearing for tracks is taken from Figure 3.9, and the particle identification efficiencies for photons, electrons and muons are taken from Figures 10.6, 10.7 and 10.8. The electromagnetic and hadronic calorimeter energy resolution are taken to be

\begin{eqnarray}
\frac{\sigma_{E}^{ecal}}{E} & = & 0.010 \oplus \frac{0.170}{\sqrt{E}} \\
\frac{\sigma_{E}^{hcal}}{E} & = & 0.094 \oplus \frac{0.559}{\sqrt{E}} 
\end{eqnarray}

\noindent as they are in the DBD Section 4.2.2 and Figure 4.15, respectively. Calorimeter tower segmentation is defined, but the nominal DSiD card uses energy flow objects rather than towers.

In the nominal DSiD card, isolation of electrons, photons and muons is not required at the detector level since isolation can be imposed after after simulation at the analysis level. It should be noted that care should be taken with the isolation since requiring it can have an enormous impact on the performance of particle identification. If isolation is not required, the jetfinding performance can be degraded. We recommend using isolation only if the primary focus of the simulation is jetfinding and not particle identification. The nominal DSiD card uses the anti $k_t$ algorithm with $\Delta R=1.0$.

For the $b$-tag, the nominal DSiD card assumes 70\% efficiency for $b$-jets, 3\% efficiency for $c$-jets, and 0.7\% efficiency for light jets. These values are taken from two points in Figure 10.9 of the DBD. Other $b$-tag operating points from that Figure can also be specified. For the $\tau$-tag, no performance is given in the DBD, so a conservative 40\% efficiency for $\tau$-jets with 0.1\% for all others are specified.

\section{Performance Validation}

\begin{figure}[p]
\begin{center}
\includegraphics[width=0.49\textwidth]{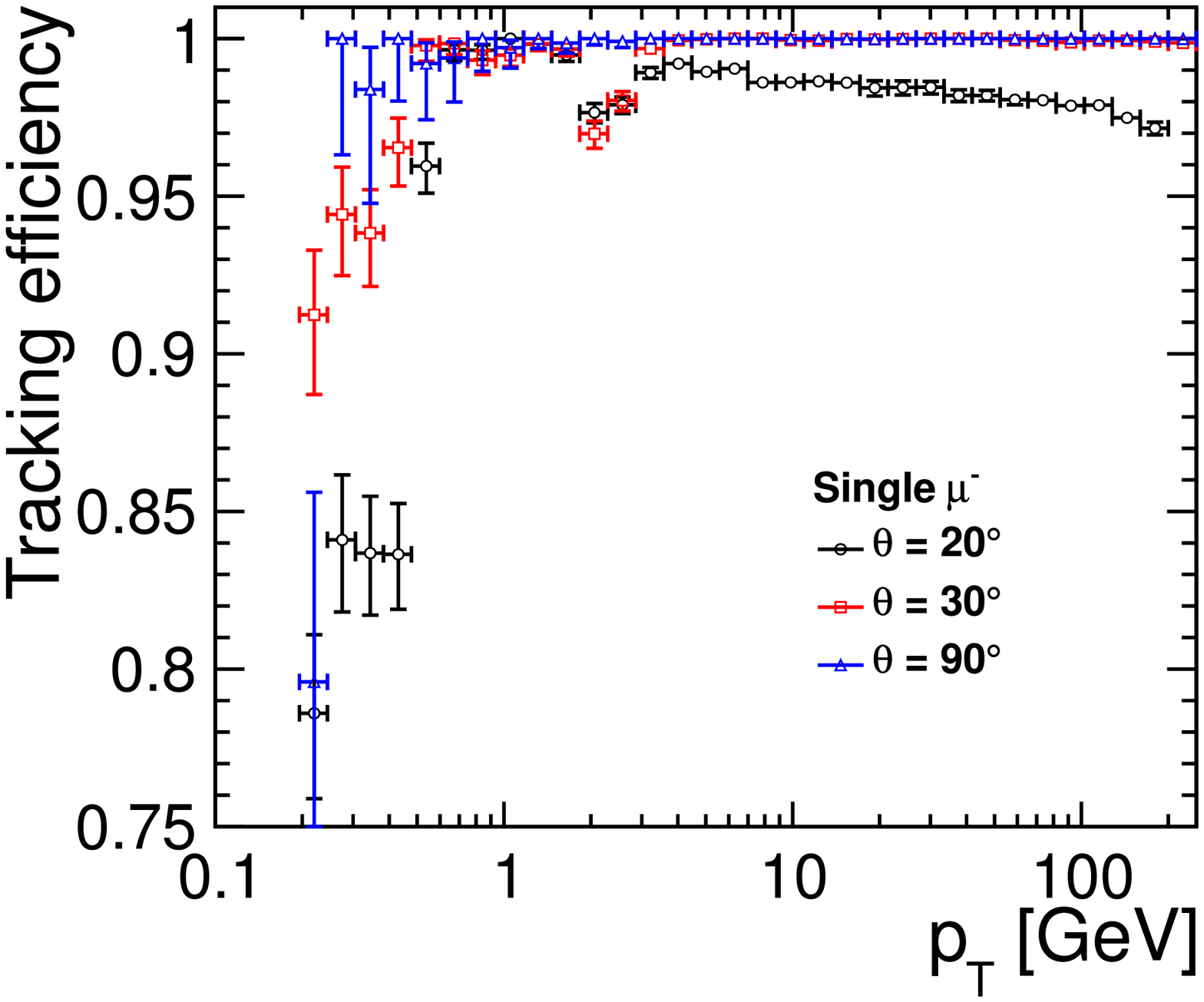}
\includegraphics[width=0.49\textwidth]{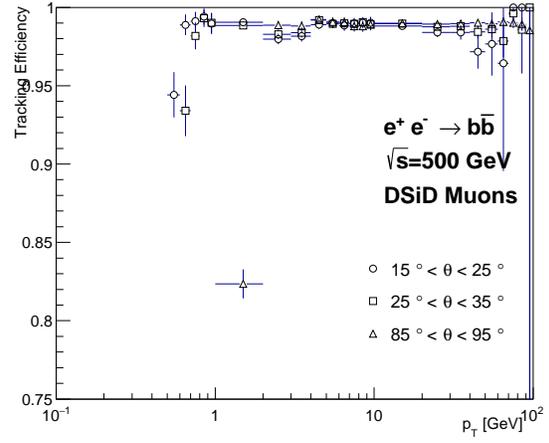}
\includegraphics[width=0.49\textwidth]{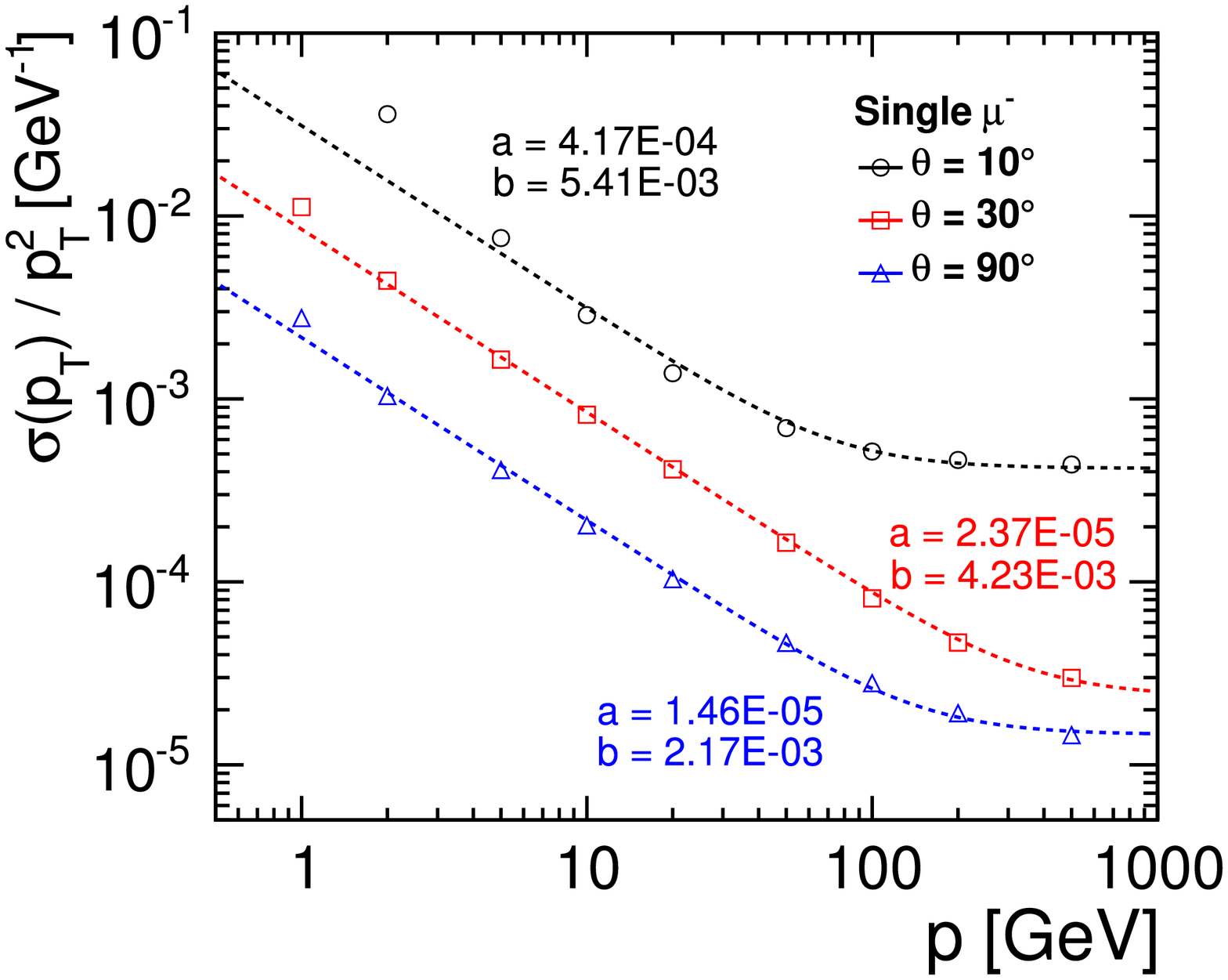}
\includegraphics[width=0.49\textwidth]{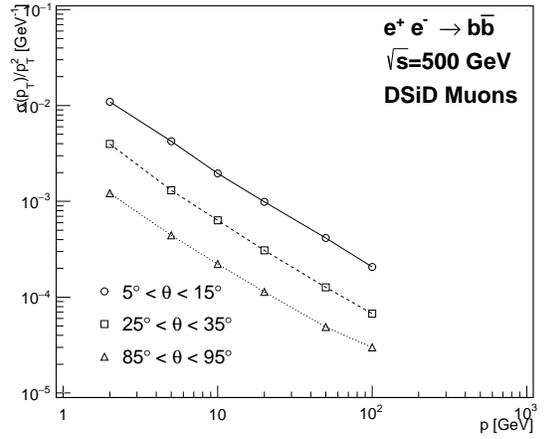}
\caption{Tracking efficiency (top) and resolution (bottom). At left, DBD Figures 3.5 (top) and 3.9 (bottom). At right, results of this study.}
\label{fig:tracking}
\end{center}
\end{figure}

\begin{figure}[p]
\begin{center}
\includegraphics[width=0.49\textwidth]{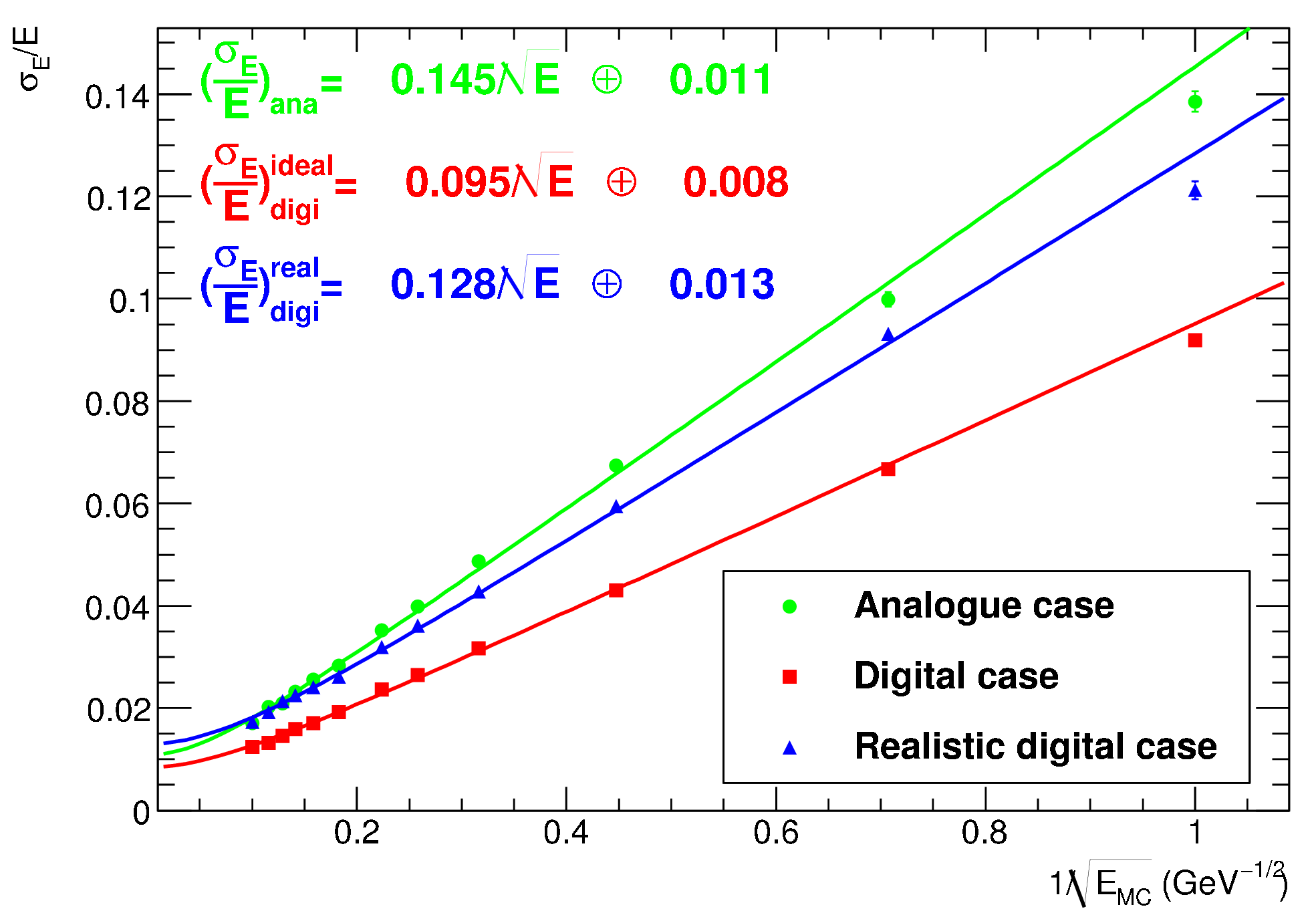}
\includegraphics[width=0.49\textwidth]{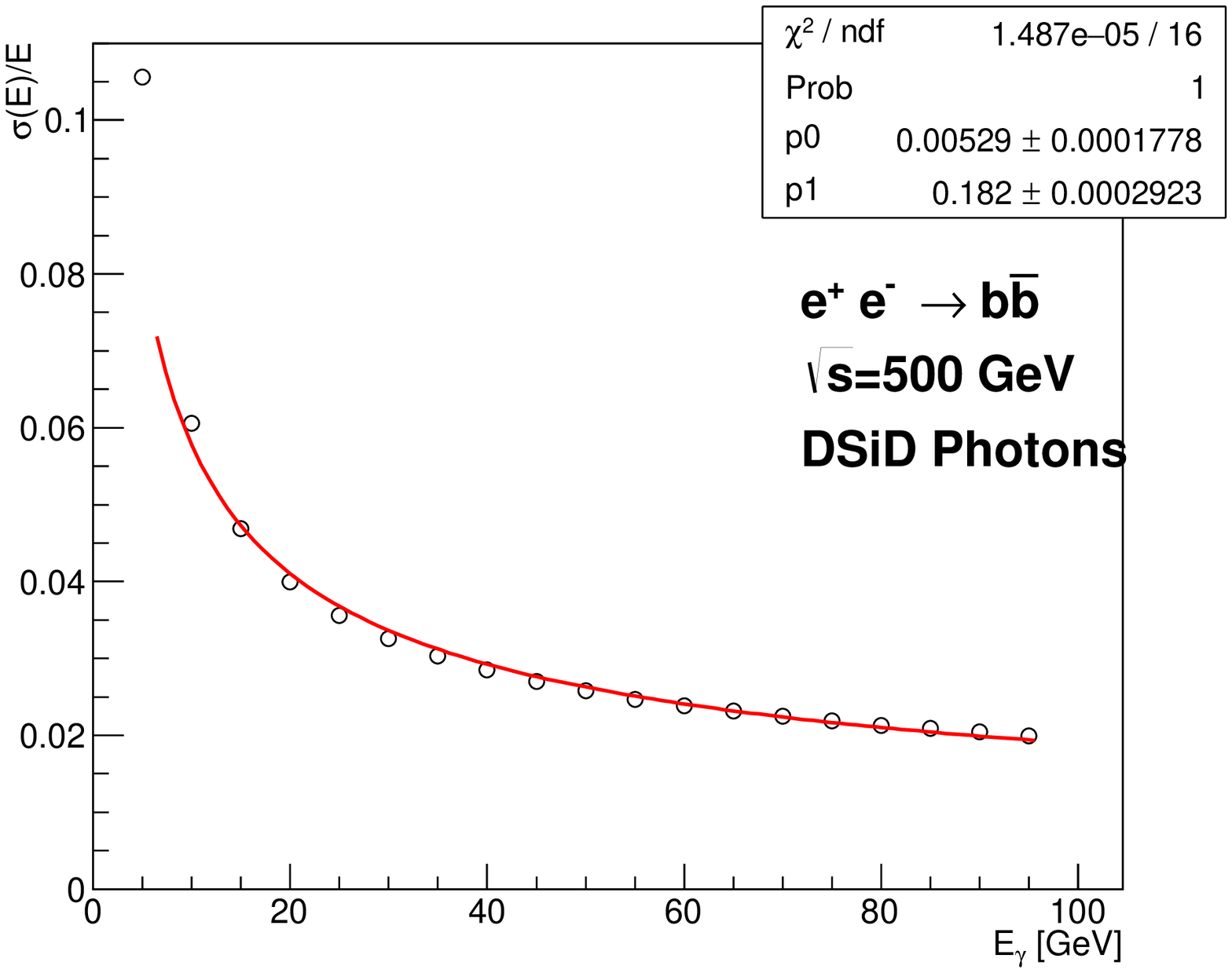}
\includegraphics[width=0.49\textwidth]{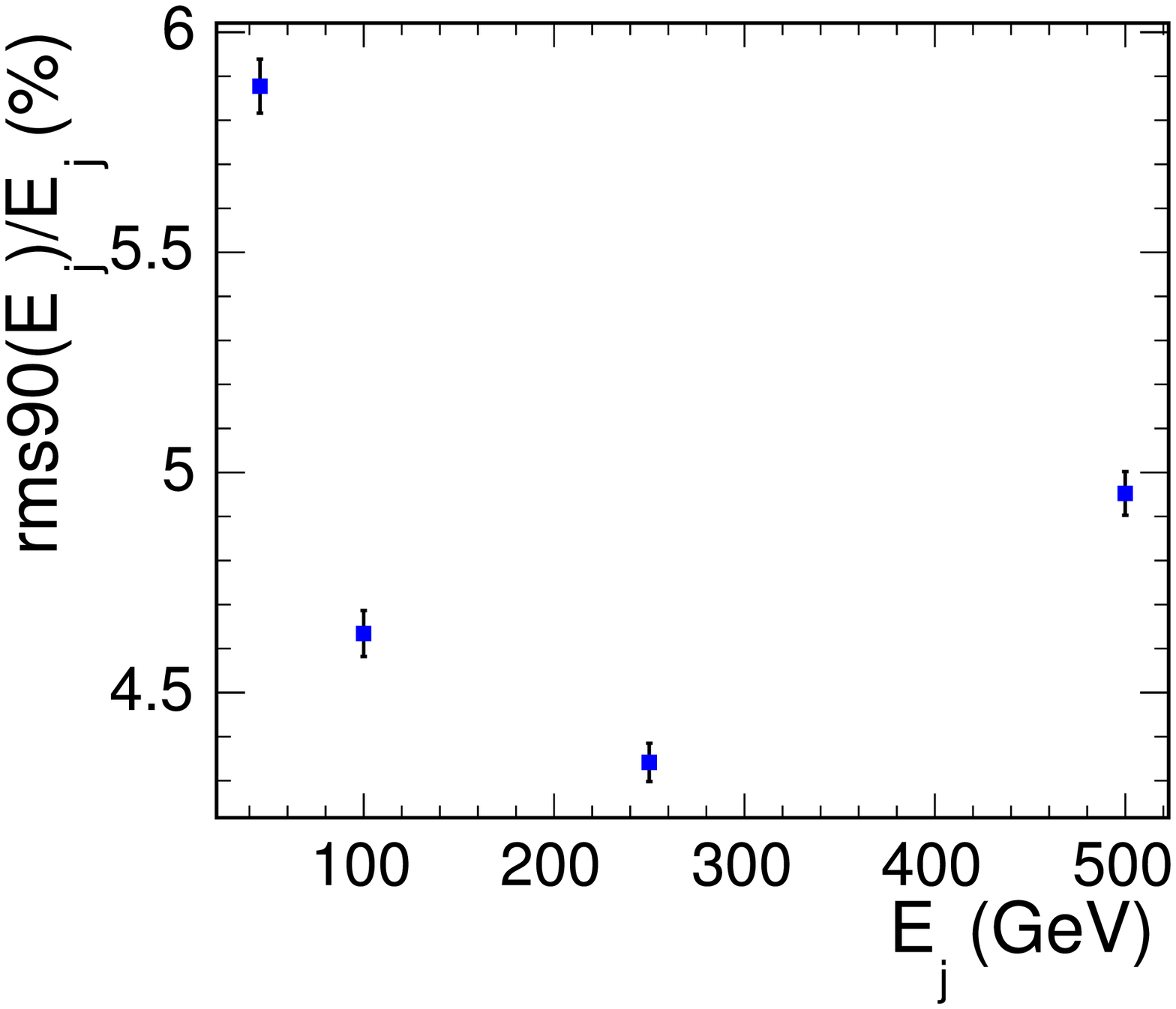}
\includegraphics[width=0.49\textwidth]{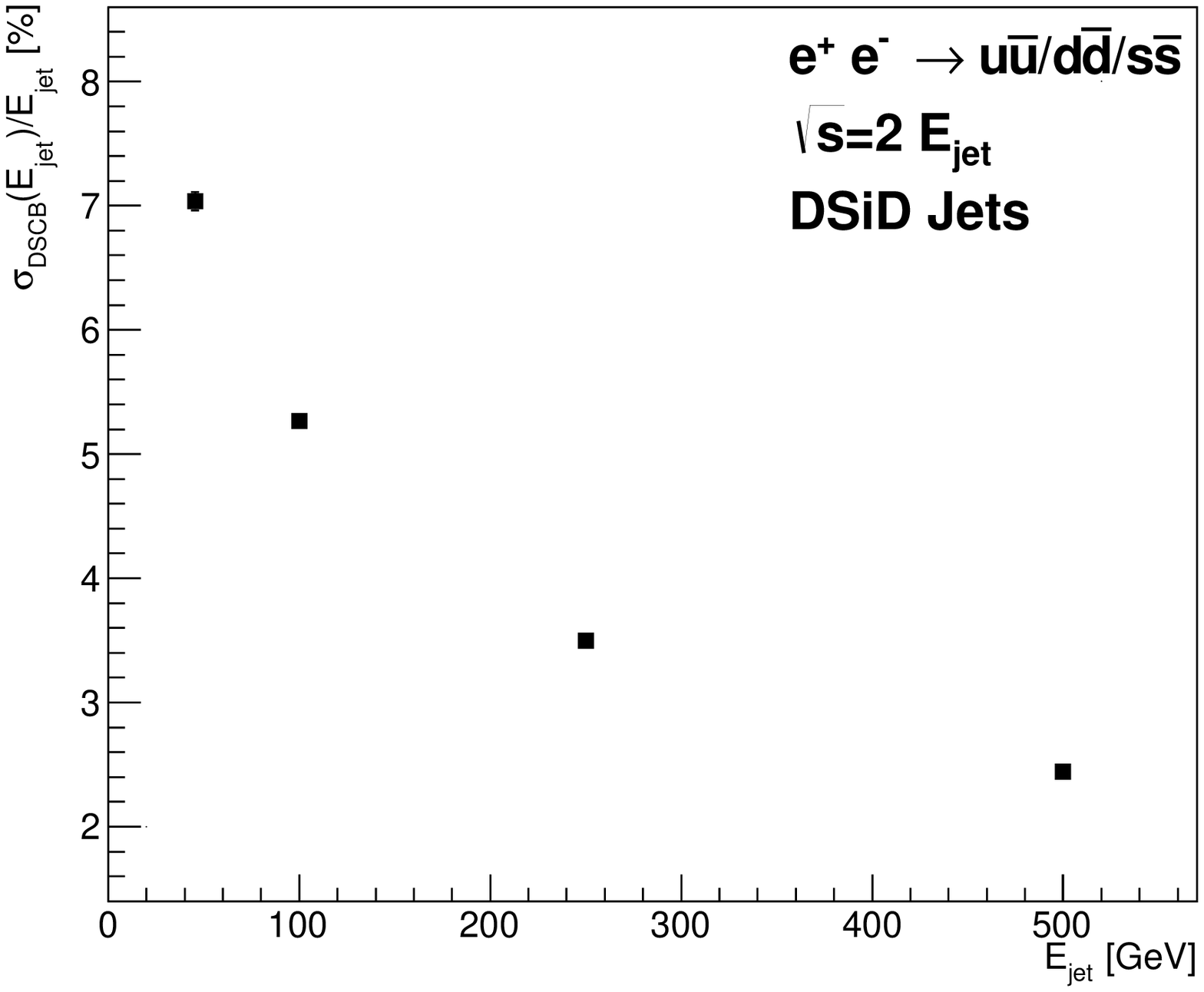}
\caption{Photon energy resolution (top) and jet energy resolution (bottom). At left, DBD Figures 4.6 (top) and 10.11 (bottom). At right, results of this study. The photon energy resolution fit is to a model $\sigma_{E}^{ecal}/E  =  p_0 \oplus p_1/\sqrt{E}$ with free parameters $p_0$ and $p_1$. Figure 4.6 shows the expected performance of the MAPS calorimeter option, not the nominal SiD electromagnetic calorimeter, and is shown here only for comparison.}
\label{fig:ecal}
\end{center}
\end{figure}

\begin{figure}[p]
\begin{center}
\includegraphics[width=0.49\textwidth]{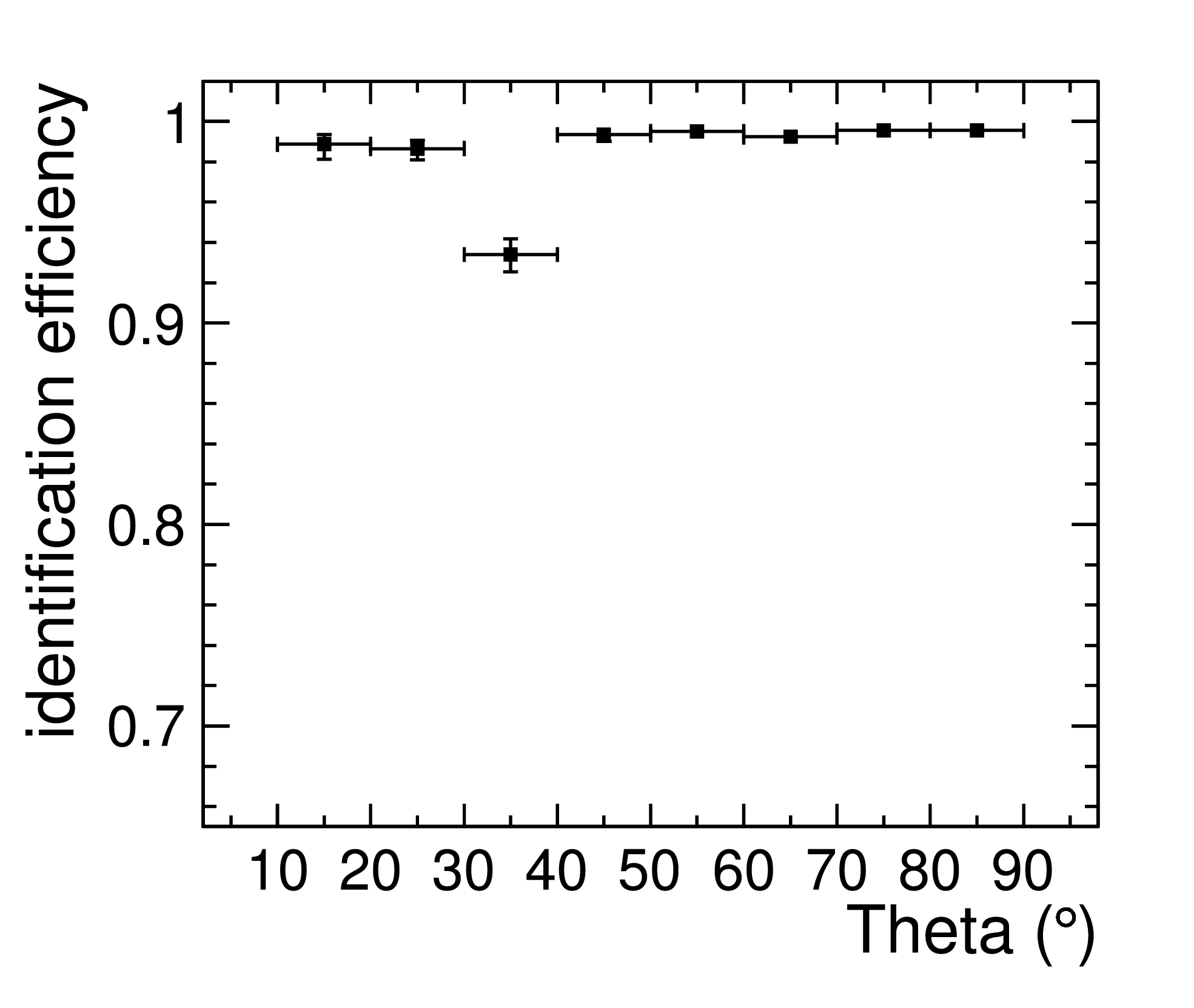}
\includegraphics[width=0.49\textwidth]{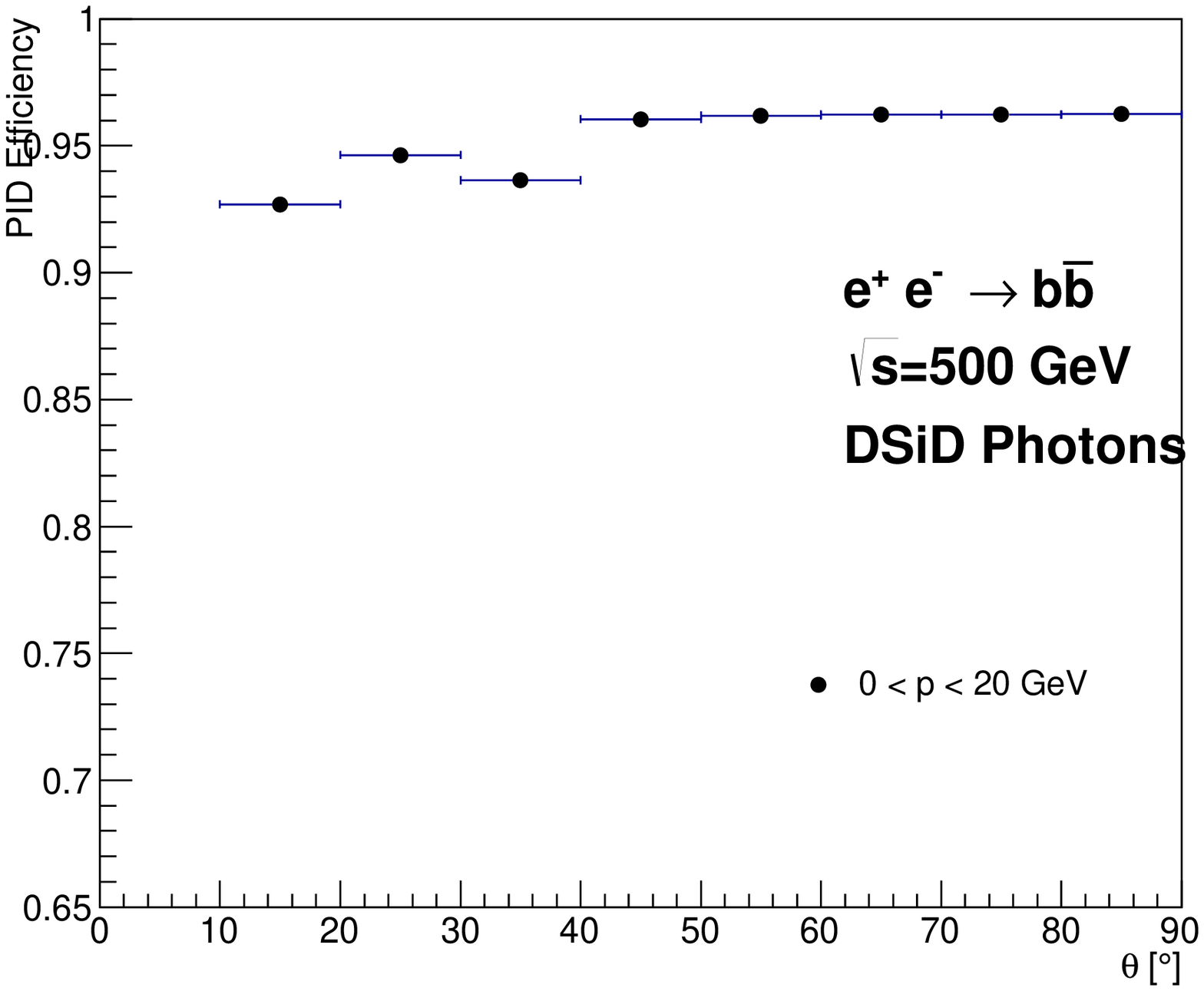}
\includegraphics[width=0.49\textwidth]{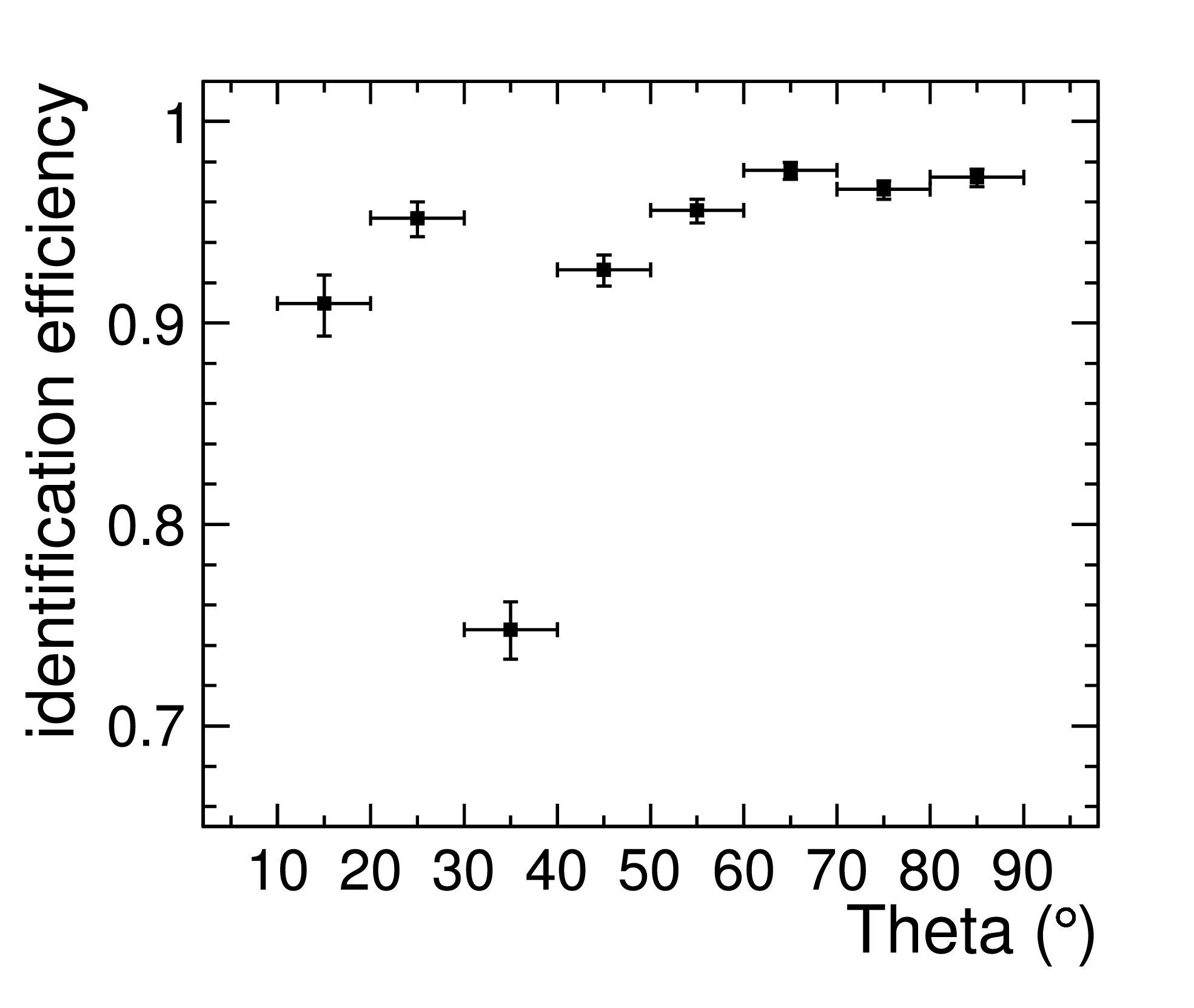}
\includegraphics[width=0.49\textwidth]{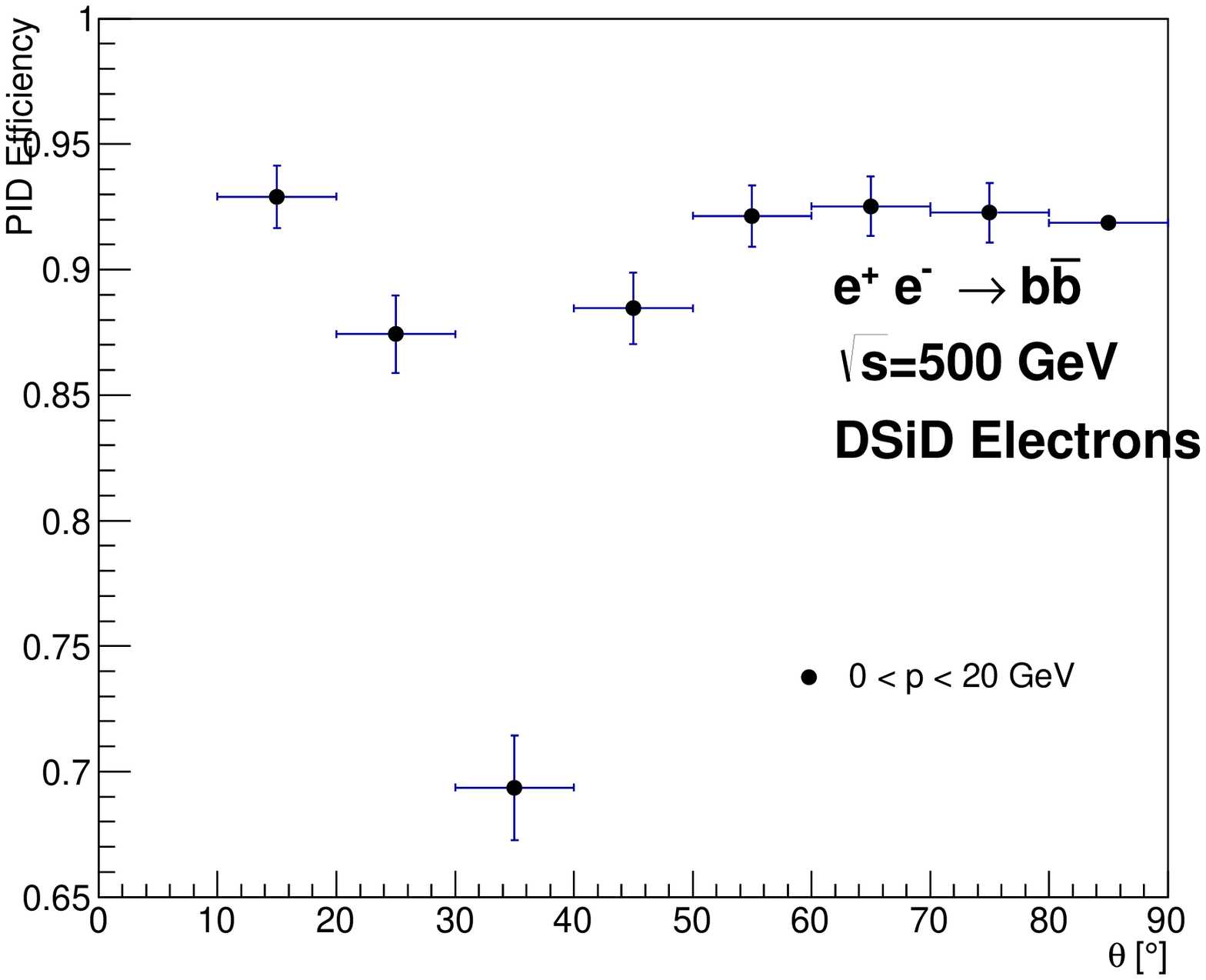}
\includegraphics[width=0.49\textwidth]{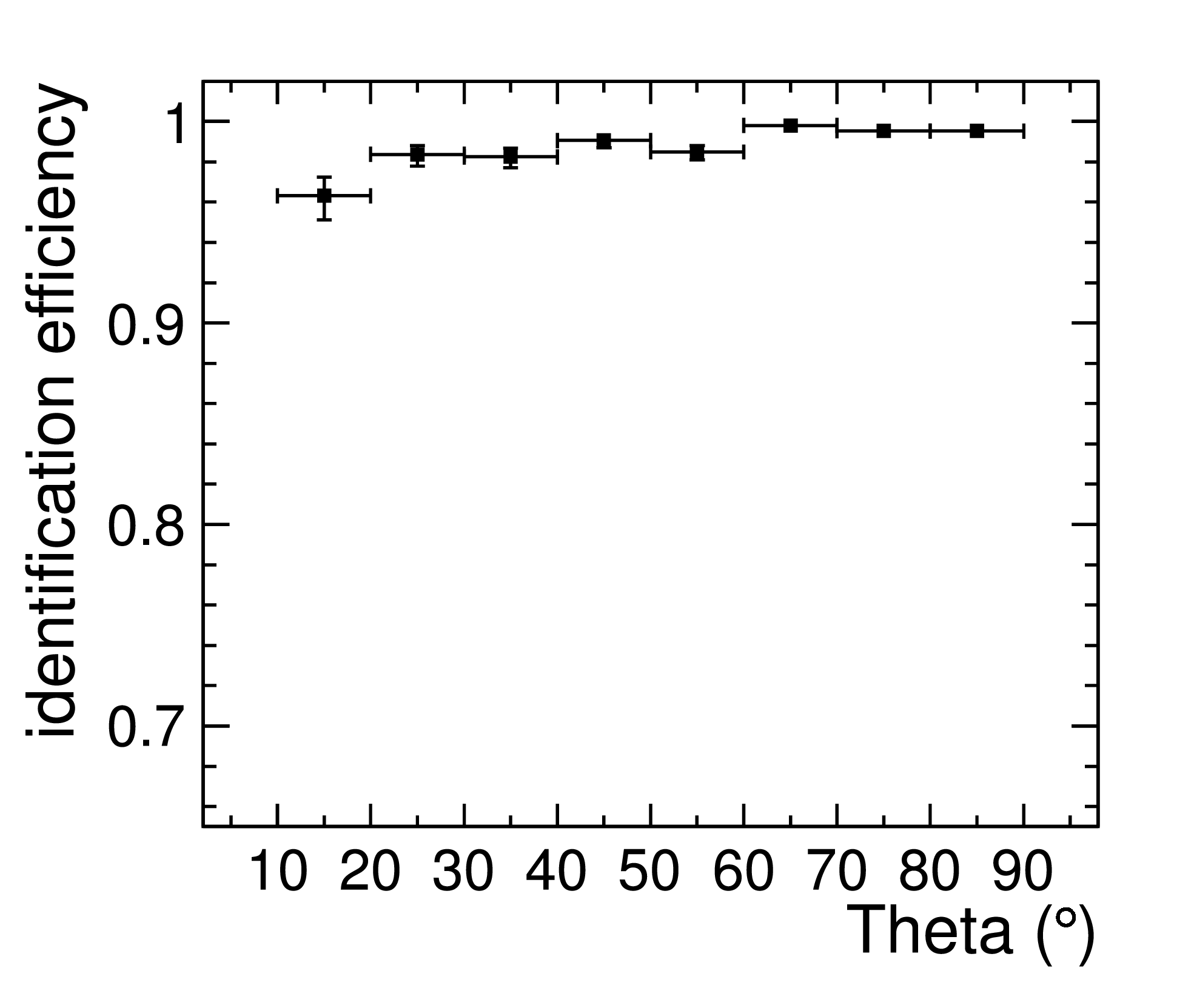}
\includegraphics[width=0.49\textwidth]{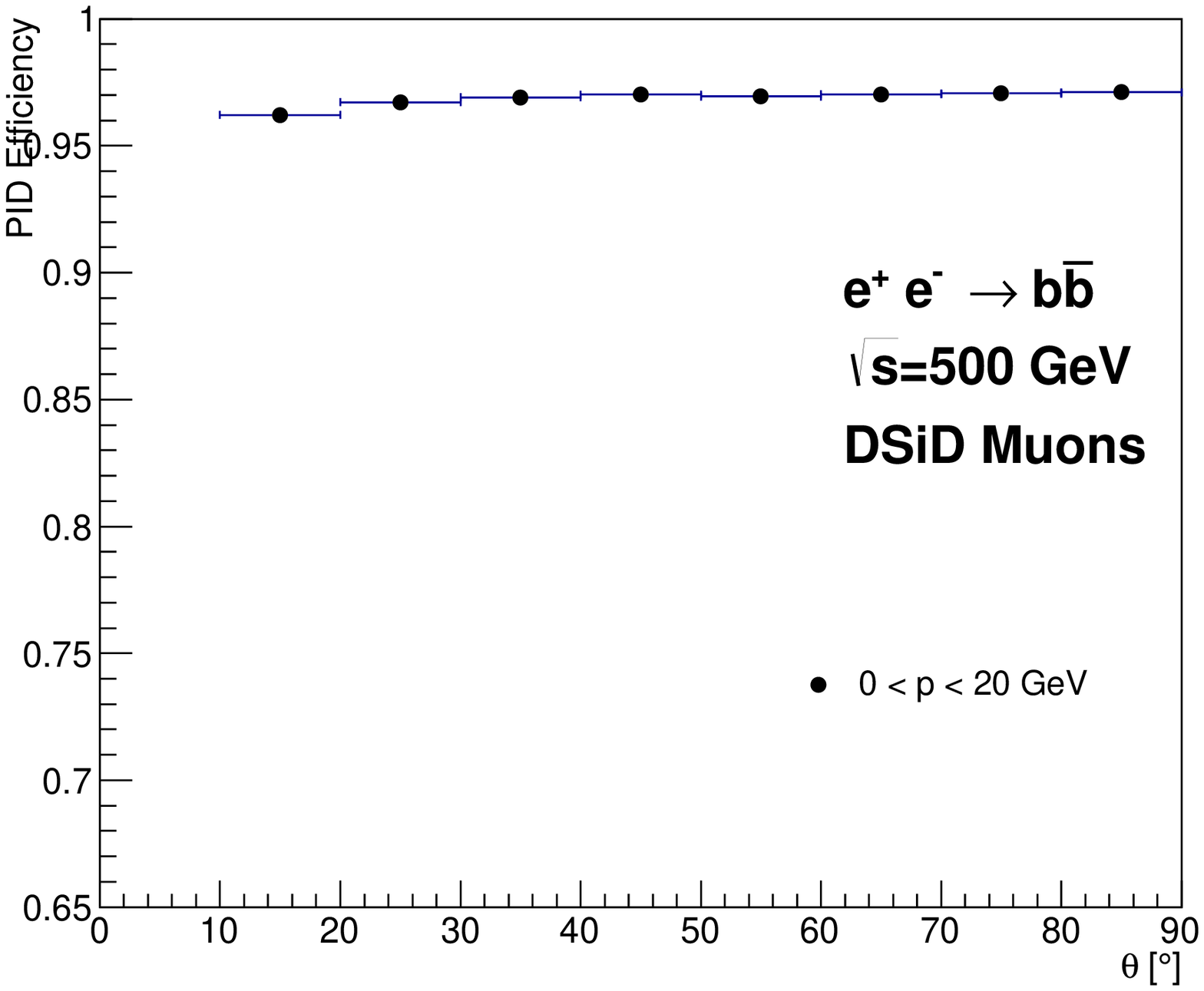}
\caption{Particle identification efficiency for photons (top), electrons (middle) and muons (bottom). At left are Figures 10.6, 10.7 and 10.8 of the DBD, which consider objects with momentum $p=10$~GeV objects. At right are the results of this study, which consider objects with momentum $0 < p < 20$~GeV. }
\label{fig:pid}
\end{center}
\end{figure}

The SiD DBD performance characteristics documented in \cite{Behnke:2013lya} were typically measured with monoenergetic or fixed-angle particles. In order to ensure that these performance characteristics can be recovered using fast simulation in a challenging event environment  with many particles and a wide spectrum of energy and momentum, we generate $4 \times 10^6$ events with MG5\_aMCv2.3.3 \cite{Alwall:2014hca}, specifying $e^+ e^- \rightarrow b \bar{b}$ at $\sqrt{s}=500$~GeV and StdHep output format. In order to measure the $b$-tag mistag rates for charm and light jets, we have also generated $10^6$ $e^+ e^- \rightarrow c \bar{c}$ events and $10^6$ $e^+ e^- \rightarrow u \bar{u}, d \bar{d}, s \bar{s}$ events. In order to study jet energy resolution, we have generated $10^6$ light dijet events $e^+ e^- \rightarrow  u \bar{u}/d \bar{d}/s \bar{s}$ events at $\sqrt{s}=91,200,500,1000$~GeV. We then use Delphes3.3.1 for detector simulation on these events.

For tracking efficiency and momentum resolution from the DBD and as measured in this study, see Figure \ref{fig:tracking}. The DBD studies use single muons with fixed polar angle $\theta=10^{\circ},20^{\circ},30^{\circ},90^{\circ}$. In this study we take muons in a range within $5^{\circ}$ of these choices for $\theta$. The performance in the complex $e^+ e^- \rightarrow b \bar{b}$ event environment matches the DBD performance well, with any differences likely due to continuous $\theta$ ranges or underpopulated phase space.

For photon and jet energy resolution, see Figure \ref{fig:ecal}. No isolation is imposed for the photon energy resolution analysis, but isolation for electrons and photons is required for the jet energy resolution analysis. Fitting the photon energy resolution distribution to  $\sigma_{E}^{ecal}/E  =  p_0 \oplus p_1/\sqrt{E}$ yields values close to, though not identical to the DBD values. The small differences are likely attributable to the low $E_{\gamma}$ and high $E_{\gamma}$ regions, which are overpopulated in the low and underpopulated in the high. 
For the jet energy resolution analysis, jets are reconstructed with the anti$k_T$ algorithm with $\Delta R=1.0$ using isolated energy flow objects. The isolation requires a maximum of 12\% (25\%) relative $E_T$ ($p_T$) in a cone of radius $\Delta R=0.5$ around electrons and photons (muons). For non-isolated objects, the jet energy resolution performance is reduced. For this study, the energy resolution is taken from the width of the core Gaussian in a double sided Crystal Ball (DSCB) fit, while for the DBD study it is taken to be the RMS of the smallest interval containing 90\% of jets. For $E_{jet}=45,100$~GeV, the energy resolution with Delphes is about 1\% worse than for the DBD jet energy resolution while it is better for $E_{jet}=250,500$~GeV. The jet energy resolution at low energy is understood since no optimization for DSiD over isolation, jet algorithm, or jet parameters has been made in this study. At high jet energy, Delphes does not account for energy leakage while the full simulation DBD result does.

For electron, photon and muon particle identification efficiency, see Figure \ref{fig:pid}. Generator muons are considered identified if the reconstructed muon points to a parent generator muon with the identical energy to the true muon energy. For the electron and photon particle identification analysis, we first reject generator electrons which lose energy to bremsstrahlung and photons which convert. For photons, conversions are first vetoed by considering only generator photons with no electrons reconstructed in a radius of $\Delta R=0.3$ around the photon. To be considered identified, a photon must be reconstructed within $\Delta R=0.3$ of the generator photon and satisfy a loose requirement on the reconstructed photon energy, requiring it to be within $5\sigma$ of the true energy. Similarly electrons losing energy to bremsstrahlung are vetoed by considering only generator electrons with no reconstructed photons in a radius of $\Delta R=0.3$ around the electron. To be considered identified, the electron must be reconstructed within $\Delta R=0.3$ of the generator electron and satisfy a loose requirement on the reconstructed electron energy, requiring it to be within $5\sigma$ of the true energy.

\begin{figure}[t]
\begin{center}
\includegraphics[width=0.49\textwidth]{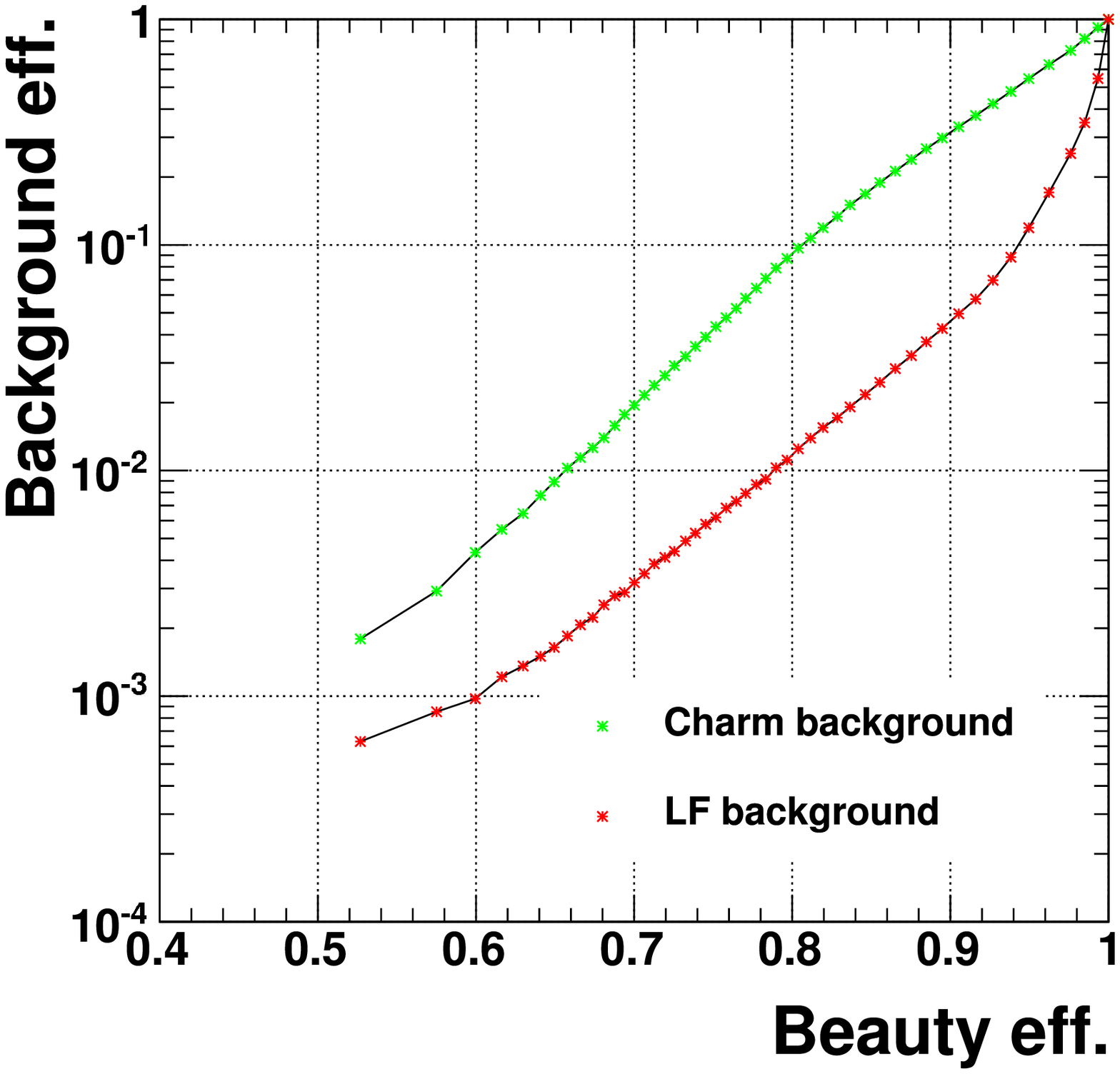}
\includegraphics[width=0.49\textwidth]{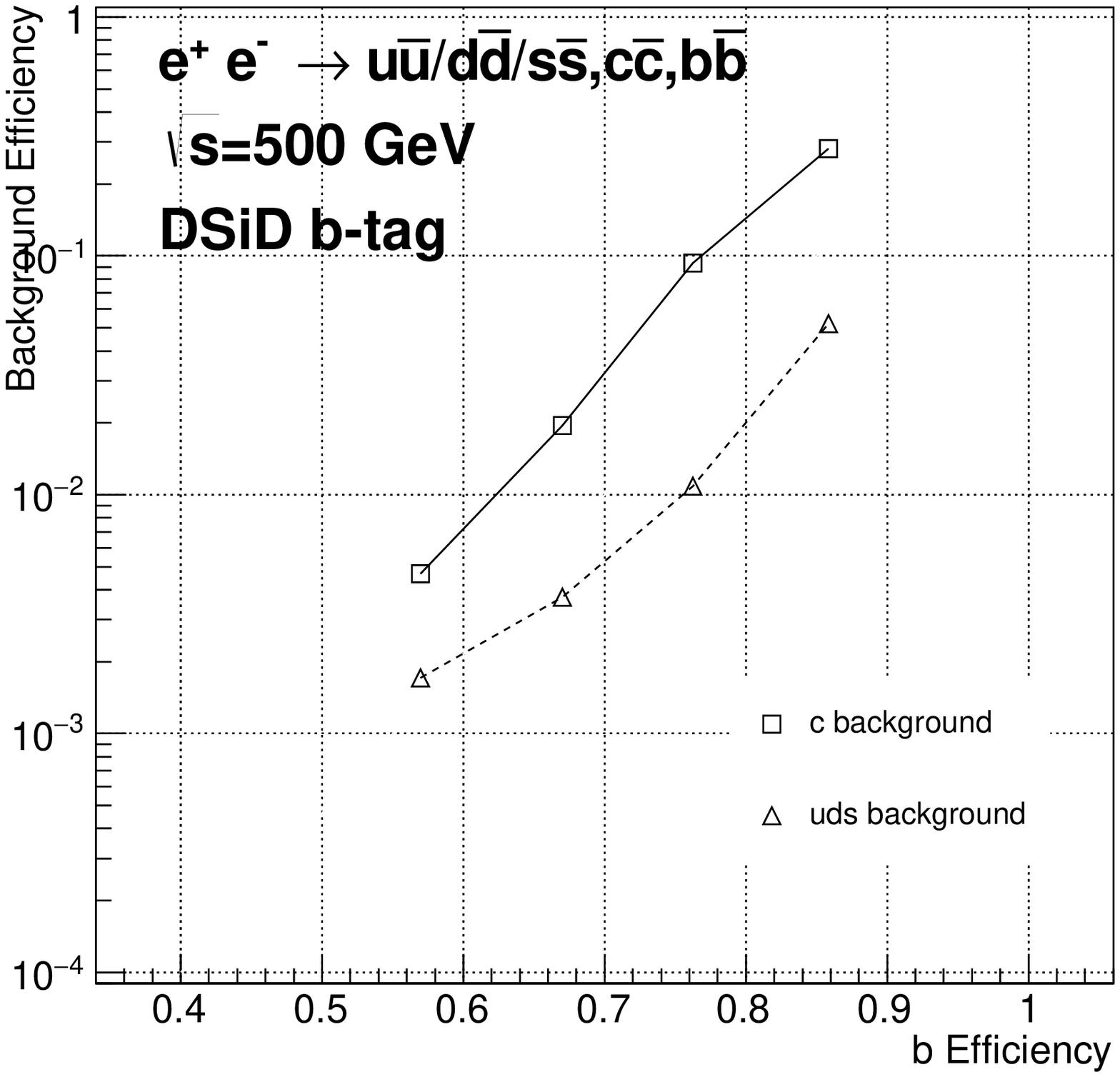}
\caption{Efficiencies for $c$- and $uds$-jet versus the $b$-jet efficiency. At left is the DBD Figure 10.9, at right is the result of this study. The $b$-tag efficiencies in DSiD are set to 60/0.4/0.1\%, 70/2/0.3\%, 80/10/1\% and 90/30/5\% for $b$/$c$/$uds$, respectively.}
\label{fig:btag}
\end{center}
\end{figure}

For the $c$- and $uds$-jet mistagging rates for the four efficiency operating points considered (60\%, 70\% 80\% and 90\%), see Figure \ref{fig:btag}. We evaluate the flavor tagging performance with DSiD by measuring the efficiency to identify reconstructed $b$-, $c$- and $uds$-jets for the nominal DSiD tagging performance as well as three other operating points described by 60\%, 80\% and 90\% $b$-tag efficiency. Events with ISR or FSR are rejected by requiring exactly two reconstructed jets.

\section{Physics Example}

To illustrate a new physics use case for DSiD, we describe a brief study of neutralino pair production at the $\sqrt{s}=500$~GeV ILC. Backgrounds are not evaluated, so the signal selection is not optimized.

The Next-to-Minimal Supersymmetric Standard Model (NMSSM) is an attractive version of Supersymmetry \cite{Martin:1997ns} which solves the $\mu$-term problem of the Minimal Supersymmetric Standard Model (MSSM) by introducing a Higgs singlet in addition to the two Higgs doublets of the MSSM \cite{Ellwanger:2009dp,Maniatis:2009re}. The NMSSM phenomenology includes seven Higgs bosons $a_1, a_2, h_1, h_2, h_3, H^+, H^-$, five neutralinos $\chi_1, \chi_2, \chi_3, \chi_4, \chi_5$ and four charginos $\chi_{1}^+, \chi_{1}^-, \chi_{2}^+, \chi_{2}^-$. For references to other phenomenological studies of NMSSM neutralino pair production at linear colliders, see \cite{Ellwanger:2009dp}.

We assume the NMSSM benchmark point $h_{60}$ described in \cite{refId0}. In $h_{60}$, pair production of all five neutralinos and all four charginos is accessible at the $\sqrt{s}=500$~GeV ILC. Moreover, through neutralino cascade decays lighter Higgs bosons $a_1$ and $h_1$ are accessible. The $125$~GeV Higgs boson discovered at the Large Hadron Collider (LHC)  \cite{Aad:2012tfa,Chatrchyan:2012ufa} is identified in $h_{60}$ as the dominantly doublet $h_2$, while the lighter $h_1$ and $a_1$ are dominantly singlet. In $h_{60}$ the singlet is largely decoupled from the doublets, allowing the singlet to avoid exclusion by LHC searches which typically assume MSSM signatures. The lightest Supersymmetric particle (LSP) in $h_{60}$ is the singlino $\chi_1$.  

In $h_{60}$, $m_{a_1} \approx 10$~GeV, $m_{h_1} \approx 56$~GeV and $m_{\chi_1} \approx 58$~GeV. The dominant decay of the $\chi_3$ is $\chi_3 \rightarrow h_1 \chi_1$. The dominant decay of the singlet dominated $h_1$ is $h_1 \rightarrow 2 a_1$, while a subdominant decay is $h_1 \rightarrow b\bar{b}$.  Since the cross section for $e^+ e^- \rightarrow \chi_3 \chi_3$ is of order 200~pb at the $\sqrt{s}=500$~GeV, this makes $e^+ e^- \rightarrow \chi_3 \chi_3 \rightarrow 2h_1 2\chi_1 \rightarrow 4a_1 2\chi_1$ a promising channel for precision measurement of the singlet Higgs sector of the NMSSM. The dominant $a_1$ decay is $a_1 \rightarrow \tau^+ \tau^-$, but the clean channel $a_1 \rightarrow \mu^+ \mu^-$ is still accessible with a branching ratio near 0.3\%.

 We assume the running scenario G-20 described in \cite{Barklow:2015tja}, namely $\int dt \mathcal{L}=5000$~fb$^{-1}$, 80\% of which is split evenly between the two $e^+ e^-$ polarization configurations $P(e^-)=+80\%,P(e^+)=-30\%$ and $P(e^-)=-80\%,P(e^+)=+30\%$. For cross section and event generation, we use MG5\_aMC@NLO \cite{Alwall:2014hca}, which features polarized beams and a complete NMSSM model that allows  $h_{60}$ spectrum and decay specification with SLHA \cite{Skands:2003cj,Allanach:2008qq}. For detector simulation we use Delphes3.3.1 \cite{deFavereau:2013fsa} with the DSiD card.  

The signal selection for the $a_1 \rightarrow \mu^+ \mu^-$ channel simply requires two oppositely charged muons with $p_{t}>20$~GeV which satisfy $\Delta R(\mu^+,\mu^-)<0.14$. Muon isolation in DSiD is modified to use a smaller cone size $\Delta R=0.1$ so that the highly collimated muons from $a_1 \rightarrow \mu^+ \mu^-$ do not fail the isolation requirement. For the $h_1 \rightarrow b\bar{b}$ signal selection, exactly three or exactly four jets are required, of which exactly two are required to be $b$-tagged. The two $b$ jets are required to satisfy $\Delta R(b,b)<\frac{\pi}{2}$. Finally, electrons and muons are vetoed to reduce the impact of semileptonic $b$ and $c$ decays to neutrinos on the $h_1$ mass reconstruction. 

\begin{figure}[t]
\begin{center}
\includegraphics[width=0.49\textwidth]{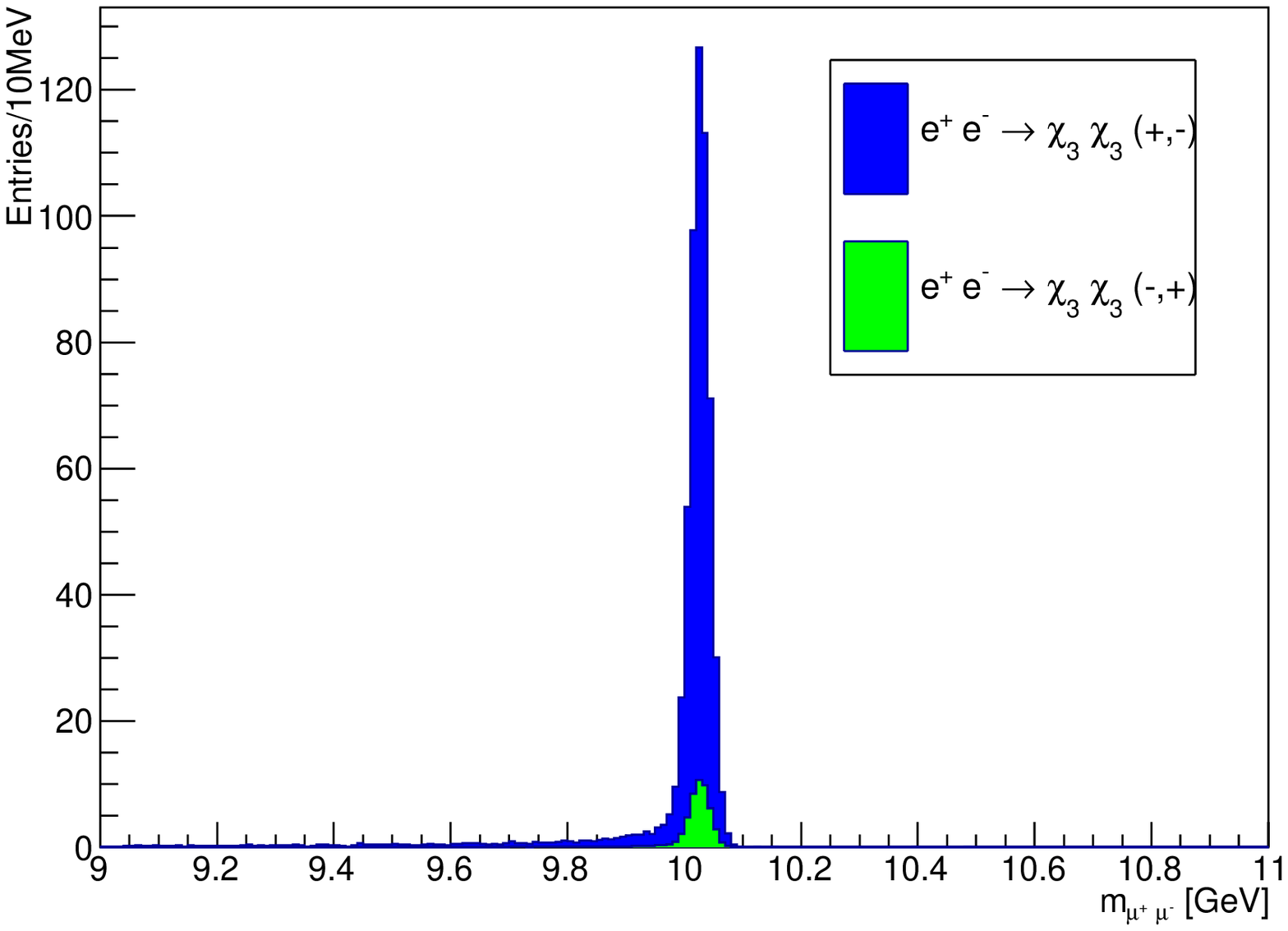}
\includegraphics[width=0.49\textwidth]{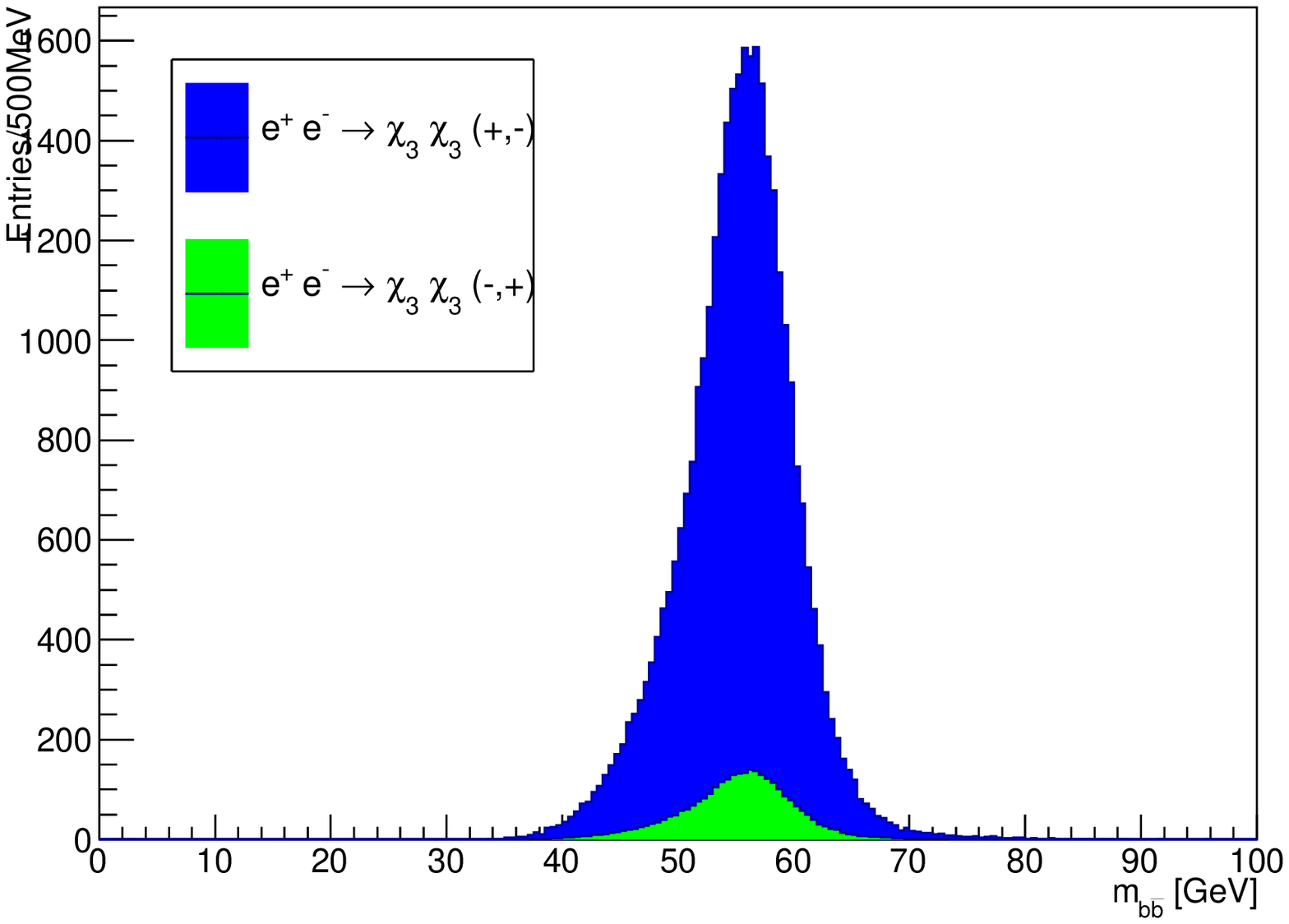}
\caption{Reconstructed $a_1 \rightarrow \mu^+ \mu^-$ (left) and $h_1 \rightarrow b \bar{b}$ (right) in $e^+ e^- \rightarrow \chi_3 \chi_3$ events at $\sqrt{s}=500$~GeV. We assume 2000fb$^{-1}$ integrated luminosity for each beam polarization configuration  $P(e^+)=+80\%,P(e^-)=-30\%$ (dark) and $P(e^+)=-80\%,P(e^-)=+30\%$ (light). The signal selection has not been optimized. The non-Gaussian structure in $h_1 \rightarrow b \bar{b}$ is due to unmeasured energy from neutrinos in semileptonic meson decays.}
\label{fig:h60}
\end{center}
\end{figure}

For the reconstructed $a_1$ and $h_1$, see Figure~\ref{fig:h60}. The non-Gaussian structure evident in the $h_1 \rightarrow b\bar{b}$ distribution is due to unmeasured neutrino energy in semileptonic meson decays. Such energy can be partially recovered with track vertexing and the $p_T$-corrected mass. A flat 5\% jet energy scale correction has been applied.

\section{Conclusion}

We have described DSiD, a fast simulation Delphes detector modeled on the SiD detector. The DSiD card parameterizes the SiD full simulation performance results documented in the ILC TDR. We then carry out validation studies to confirm that the card can reproduce the SiD performance in a complex event environment $e^+ e^- \rightarrow b \bar{b}$ at $\sqrt{s}=500$~GeV. We conclude that, with some caveats, DSiD performance is consistent with the full simulation SiD DBD study.

We recommend Delphes for ILC phenomenology studies. As an example use case, we reconstruct $\chi_3 \rightarrow \chi_1 h_1$ in pair neutralino production at the $\sqrt{s}=500$~GeV ILC with $h_{1} \rightarrow b\bar{b}$ and $h_1 \rightarrow 2a_1$ with at least one $a_1 \rightarrow \mu^+ \mu^-$. Evaluation and tuning of the DSiD card will be ongoing at \texttt{dsid.hepforge.org}, and we invite feedback.

\begin{center}\textbf{Acknowledgements}\end{center} The author thanks the SiD detector community for encouraging feedback and the Alder Institute for High Energy Physics for financial support. 

\bibliography{paper}

\end{document}